\documentclass[conference]{IEEEtran}

\IEEEoverridecommandlockouts

\usepackage{cite}
\usepackage{graphicx}
\usepackage{float}
\usepackage{mathrsfs}
\usepackage{amsfonts}
\usepackage{color}
\usepackage{multirow}

\usepackage{amsmath,amssymb,amsfonts} 
\usepackage{algorithmic} 			
\usepackage{textcomp} 				
\usepackage{xcolor} 				

\usepackage{booktabs} 
\usepackage{verbatim}

\usepackage{subfigure}
\usepackage{url}
\usepackage{ulem}

\ifCLASSINFOpdf
\else
\fi
\begin{document}

\title{Deep Learning Based MAC via Joint Channel Access and Rate Adaptation}

\author{Jiantao Xin$^\star$, Wensen Xu$^\star$, Yucheng Cai$^\star$, Taotao Wang$^\star$, Shengli Zhang$^\star$, Peng Liu$^\dagger$, Ziyang Guo$^\dagger$, Jiajun Luo$^\dagger$ \\
$^\star$College of Electronics and Information Engineering, Shenzhen University, Shenzhen, China\\
$^\dagger$Wireless Technology Lab, 2012 Labs, Huawei, China\\
Email:
\{xinjiantao, xuwensen, caiyucheng, ttwang, zsl\}@szu.edu.cn,\\ 
\{jeremy.liupeng, guoziyang, luojiajun4\}@huawei.com.

}

\maketitle
\begin{abstract}

The existing medium access control (MAC) protocol of Wi-Fi networks (i.e., carrier-sense multiple access with collision avoidance (CSMA/CA)) suffers from poor performance in dense deployments due to the increasing number of collisions and long average backoff time in such scenarios. To tackle this issue, we propose an intelligent wireless MAC protocol based on deep learning (DL), referred to as DL-MAC, which significantly improves the spectrum efficiency of Wi-Fi networks. The goal of DL-MAC is to enable not only intelligent channel access but also intelligent rate adaptation. To achieve this goal, we design a deep neural network (DNN) that takes the historical received signal strength indications (RSSIs) as inputs and outputs joint channel access and rate adaptation decision. Notably, the proposed DL-MAC takes the constraints of practical applications into account and the DL-MAC is evaluated using the real wireless data sampled from the actual environments on the 2.4GHz frequency band. The experimental results show that our DL-MAC can achieve around 86\% performance of the global optimal MAC, and around the double performance of the traditional Wi-Fi MAC in the environments of our lab and the Shenzhen Baoan International Airport departure hall. 

\end{abstract}


\IEEEpeerreviewmaketitle


\section{Introduction}
\label{sect:Introduction}
The rapid growing of short-range communication techniques, such as Wi-Fi, Bluetooth, ZigBee, MulteFire and New Radio Unlicensed (NR-U), poses huge challenges to the user experiences due to the coexistence of dense devices on the unmannered unlicensed bands. In such crowed and heterogeneous wireless environments, the conventional design paradigm of medium access control (MAC) protocols, which were originally targeted for homogeneous networks where all the nodes follow a same rule,  has been inefficient. For example, observed from \cite{chiasserini2002coexistence,jiang2017performance}, the popularly-used carrier-sense multiple access with collision avoidance (CSMA/CA) MAC protocol of Wi-Fi networks fails to coexist with other MAC protocols, even though with  the simplest Time Division Multiple Access (TDMA) MAC.

Another issue of traditional MAC design paradigm is that the algorithm associated with some certain function is designed independently of other functions. Taking the rate adaptation concerned in this work as an example, the algorithms of the rate adaptation function are usually designed to adapt to the path loss of wireless channels, and they are working independently of the function of channel access protocol (e.g., CSMA/CA in Wi-Fi). If we shift the paradigm and devise a novel MAC algorithm in a holistic way that joins channel access and rate adaptation, much performance gain could be achieved.

\begin{figure}[!t]
	\centering
	\includegraphics[width=3.0in]{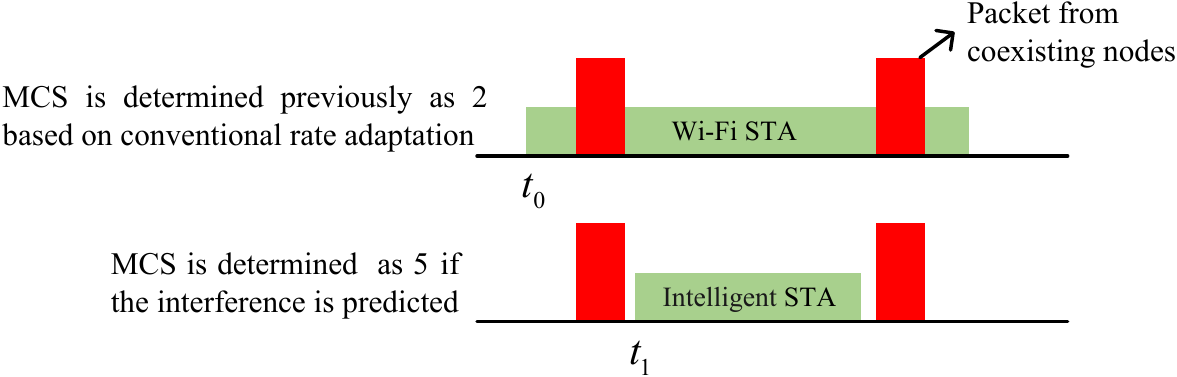}
	\caption{A toy example that shows the benefit from joint channel access and rate adaptation.}
	\label{fig_illustration}
	\vspace{-5mm}
\end{figure}

To illustrate this point, we present a toy example in Fig.~\ref{fig_illustration}, where the green packets are transmitted by the nodes of interest and the red packets are transmitted by other coexisting nodes. Fig.~\ref{fig_illustration} shows the coexistence behavior when the node of interest is a Wi-Fi station (STA) or it is an intelligent STA that uses a learning-based MAC protocol. For the Wi-Fi STA, the channel access protocol is CSMA/CA and the rate adaptation algorithm is I\textsc{wl} \cite{grunblatt2019simulation}. At time $t_0$, the Wi-Fi STA determines to transmit after a period of carrier sensing until the backoff counter decreases to zero. \footnote{Note that rate adaptation and MCS selection both refer to the task of regulating transmission rate based on the wireless channel quality. Hence, we use them interchangeably in this paper.}Suppose that the I\textsc{wl} selects a modulation and coding scheme (MCS) indexed by 2 for the Wi-Fi STA to transmit. We can see that the collision occurs between the Wi-Fi STA and the coexisting node. By contrast, for the intelligent STA, the collision can be avoided by learning the wireless environment, predicting the transmit behavior of coexisting nodes, and determining when to access the channel along with a properly selected MCS. In Fig.~\ref{fig_illustration}, the intelligent STA leans to select the MCS indexed by 5, which has a higher transmission rate than MCS 2, to transmit at time $t_1$. 

As illustrated in Fig.~\ref{fig_illustration}, the key to designing an intelligent MAC algorithm is the capability of learning the transmission behavior of coexisting nodes via observing the wireless environment. This can be achieved by employing the state-of-the-art data-driven deep learning (DL) framework. Specifically, \textbf{the MAC algorithm is implemented by a deep neural network (DNN)} that takes a time sequence of channel sensing results as inputs and outputs a decision of joint channel access and rate adaptation.  Fortunately, channel sensing results are always available for free from energy detection modules commonly employed in wireless devices. Therefore, the DNN can be used to infer whether to access the channel and which transmission rate can be used based on the historical channel sensing results.

Recently, learning-based approaches were introduced to adjust the parameters of CSMA/CA in \cite{edalat2018machine,edalat2019dynamically} and enable channel access prediction in \cite{yang2018scalable,yang2019machine,yu2019deep,yu2020non}. Moreover, the authors of \cite{li2020practical,cho2021reinforcement} proposed learning-based rata adaptation approaches for wireless networks. However, it is worth noting that these works are either merely designed for channel access or just for rate adaptation attempting to improve their respective performances rather than considering them jointly. To fill this gap, we propose a learning-based MAC protocol that achieves joint channel access and rate adaptation.

An inherent question arises that how much gain can be achieved from the joint design in realistic scenarios. To answer this question, we evaluate our learning-based MAC protocol using real wireless data sampled from actual wireless environments. We sampled wireless data from Shenzhen University and Shenzhen Baoan International Airport, and open-sourced the data set on github \cite{postman511RSSIdataset}. We then design a DNN and train it using the data from \cite{postman511RSSIdataset}. To the best of our knowledge, we are the first to propose a DL-based joint channel access and rate adaptation scheme, and to evaluate its practical performance using real wireless data. The contributions of this work are summarized as follows.

i) We propose a DL-based MAC protocol, referred to as DL-MAC, aiming at determining both the channel access and rate adaptation actions. Notably, the proposed DL-MAC exploits the real wireless data captured in the actual wireless environments.
	
ii) As a benchmark for evaluating performance, we develop a global optimal MAC that has a global perspective of future channel conditions to optimize its MAC behaviors. We trained the DNN using the sampled data and experimentally compare its performance with the global optimal MAC and the traditional Wi-Fi MAC. The experimental results show that the performances of our DL-MAC can approach that of the global optimal MAC and outperform that of the CSMA/CA with traditional rate adaptation schemes	\cite{lacage2004ieee,grunblatt2019simulation}. Moreover, we perform generalization experiments for DL-MAC, i.e., we train the DNN using the data from a specific channel and test its performance over other Wi-Fi channels. And the results of the generalization experiments also show that the DL-MAC still outperforms the traditional MAC and approach the global optimal MAC.

iii) To deeply analyze the respective gains yielded by channel access and rate adaptation in the joint design, we also develop the DL-based channel access and DL-based MCS selection schemes as the benchmarks. The experimental results reveal that the performance of the joint design outperforms both the performance of DL-based channel access and DL-based MCS selection. In addition, the respective gains of MCS selection and channel access show different behavior in different environments. Specifically, on a gentle channel condition, i.e., the interference on the channel is relatively small (as in our lab at Shenzhen University), MCS selection has a higher performance gain than channel access does, thus MCS selection is of vital importance to achieve a high throughput in such wireless environments. On the contrary, on a severe channel condition, i.e., the interference on the channel is relatively strong (as in the departure hall of Shenzhen Baoan International Airport), channel access has more significant impact on the throughout improvement than the MCS selection does.

\section{System Descriptions and Preliminaries}
\label{sect:System Descriptions}
\begin{figure}[!t]
	\centering
	\includegraphics[width=2.4in]{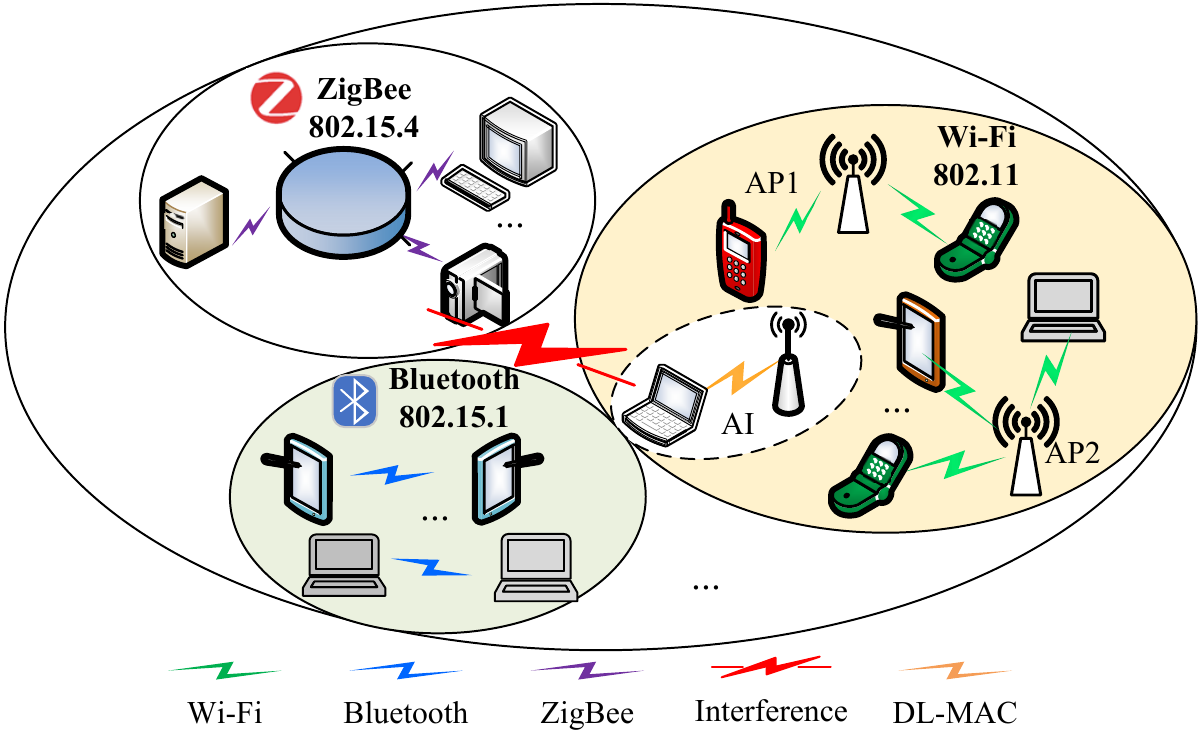}
	\caption{The wireless scenario of 2.4GHz band where devices employing different wireless protocols coexist with each other.}
	\label{fig_ISM}
	\vspace{-5mm}
\end{figure}

As shown in Fig.~\ref{fig_ISM}, we consider a Wi-Fi network working on 2.4GHz frequency band. The network consists of legacy Wi-Fi devices and \textit{AI devices}, which employ the CSMA/CA MAC protocol and the proposed DL-MAC protocol, respectively. Apart from Wi-Fi systems, there may also exist other devices operating under different wireless protocols, such as Bluetooth, ZigBee and LTE-U. These coexisting networks lead to severe interference and degrade the communication quality of Wi-Fi links. Therefore, it is urgently needed to pursue new intelligent MAC protocols which improve the spectrum efficiency. Before going into the details of the proposed DL-MAC, we first briefly introduce the CSMA/CA MAC protocol and rate adaptation mechanisms in the following.

\subsection{CSMA/CA}
The basic principle behind CSMA/CA is listen before talk \cite{hiertz2010ieee}, i.e., the node needs to listen to the channel before each packet transmission. If the channel is sensed idle for a time period of Distributed Inter-frame Spacing (DIFS), the node enters into a random backoff stage. The backoff count is decreased by one if the channel is sensed idle for a mini-slot and a transmission is allowed if the backoff count reaches zero. Once the channel is sensed occupied, the node will keep silent and listen to the channel until it is sensed idle for another DIFS period. A binary exponential backoff scheme is adopted by Wi-Fi MAC for collision avoidance, that is, doubling the average backoff count after each transmission collision. Hence, after a failed transmission, Wi-Fi STA will wait for a longer time. We can see that when to transmit is highly determined by the result (success or not) of the last transmission. Differently, channel access decision of DL-MAC is made via a DNN based on a sequence of channel sensing results.

\subsection{Rate Adaptation}
Rate adaptation is proposed to be an adaptive algorithm for channel variations in wireless environments \cite{lacage2004ieee,grunblatt2019simulation}. MCS should be tuned carefully as incorrect decisions may lead to drastic performance degradation. A widely used rate adaptation algorithm on licensed band is based on the signal-to-noise ratio (SNR), i.e., the transmitter determines the MCS based on the SNR feedback from the receiver or the estimation from the inverse link. However, on unlicensed band, SNR-based approach usually doesn't work well due to its high interference environment, since the optimal MCS depends not only on the noise power but also on the interference from coexisting nodes. In practice, most rate adaptation algorithms on unlicensed band are sampling-based \cite{grunblatt2019simulation}, which means that sampling statistical values, e.g., packet error rate (PER), throughput, from all rates and transmitting at the rate with the best performance in history. But this sacrifices some performance since the sampling-based algorithm has to explore with a certain probability at a suboptimal rate to get sufficient statistics. Differently, DL-MAC predicts the optimal MCS based on the historical channel sensing results, no need to transmit at all rates.

\section{Data Set and Processing}
Our proposed DL-MAC exploits the real wireless data sampled from the 2.4GHz frequency band to build a DNN that implements channel access and rate selection. Therefore, this section gives a brief introduction of the real sampled data set we collected and the process of processing the data set. More detailed description is elaborated on the public data set website \cite{postman511RSSIdataset}.

\subsection{Raw Data Set}
The received signal strength indications (RSSIs) as the channel sensing results are freely obtained via the energy detection modules of wireless devices. Then, we capture RSSI data on the 2.4 GHz frequency band using the spectrum analysis function of Ellisys Bluetooth Vanguard, and the public data set can be accessed in \cite{postman511RSSIdataset}. The raw RSSI dataset is a matrix of $L$ RSSI samples ($100\mu s$ sampling interval) times $N=83$ sub-bands (1MHz bandwidth) and can be represented as:
\vspace{-1mm}
\begin{equation}\small
\setlength{\arraycolsep}{1pt}
\boldsymbol{RSSI}=\left[\begin{array}{ccc}
RSSI_{0, 0}, & \cdots, & RSSI_{N, 0} \\
\vdots & \vdots & \vdots \\
RSSI_{0, L}, & \cdots, & RSSI_{N, L}
\end{array}\right] \in \mathbb{R}^{L \times N}.
\end{equation}

Note that the collected raw RSSI data contain signals of all wireless devices operating on this band.

\subsection{Data Preprocessing}
Rather than directly sending the raw data to our DL-MAC for use, we need to preprocess the raw RSSI samples in both the time and frequency domains. We perform interpolation sampling algorithms to sequentially align the mini-slot length ($9\mu s$) of CSMA/CA and the available channels (13 channels with 20MHz bandwidth) in the Wi-Fi network, respectively. Note that, in this work, DL-MAC does not include channel selection, thus, we only concern the DL-MAC operating on a specific preset channel. For simplicity, by omitting channel index, the RSSI matrix is reduced to an RSSI vector,
\begin{equation}\small
\boldsymbol{RSSI}^{\prime}=\left[RSSI_0^{\prime},\cdots, RSSI_L^{\prime}\right]^{T}.
\end{equation}

\section{Deep Learning Based MAC}
\label{sect:Proposed Deep Learning-Based Mac}
In this section, we first introduce the DNN architecture followed by how to train the DNN with the historical RSSIs. Then, we present how to use the trained DNN to implement our DL-MAC.

\subsection{DNN Model}
\begin{figure}[!t]
	\centering
	\includegraphics[width=2.4in]{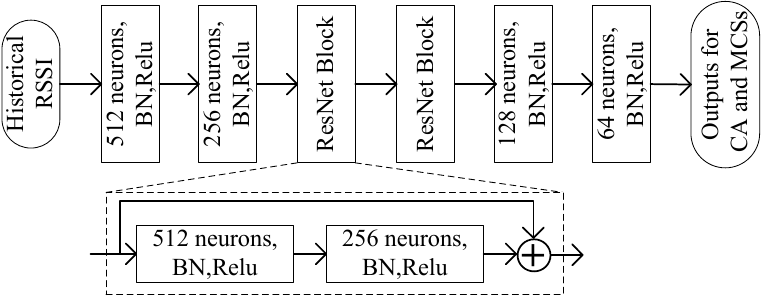}
	\vspace{-3mm}
	\caption{Architecture of DNN.}
	\label{fig_DNN_ResNet}
	\vspace{-5mm}
\end{figure}

As illustrated in Fig.~\ref{fig_DNN_ResNet}, the architecture of the DNN used in our DL-MAC is an eight-hidden-layer ResNet, with a various number of neurons in each hidden layer. Specifically, in each hidden layer, a batch normalization (BN) layer and an activation function layer are sequentially incorporated to constitute a composite hidden layer. All the hidden layers are fully connected (FC) and the first two hidden layer consists of 512 and 256 neurons, sequentially. By following two ResNet blocks, the neuron numbers of the two subsequent hidden layers decrease to 128 and 64, respectively. We adopt batch normalization before the activation function that is Rectified Linear Unit (ReLu) to guarantee neither the forward nor the backward propagation signal vanishes. For each ResNet block, two FC hidden layers and one identity shortcut connection are contained.

The input is the historical RSSIs whose length is equal to 3 times the maximum transmission opportunity (MTXOP) of a packet, that is the time required for the DL-MAC to transmit with minimum MCS. For instance, we suppose that an MTXOP is equal to 120 mini-slots. At time $t$, the input is $RSSI_{t-359}^{\prime}, RSSI_{t-358}^{\prime}, \cdots, RSSI_{t}^{\prime}$. The output layer using the SoftMax activation function is composed of 10 neurons that represent the probability of each MCS index. As shown in Table~\ref{tab_MCS}, 9 MCSs, i.e., 0 to 8, are considered. In addition, we add an MCS index of -1 to indicate that the DL-MAC decides not to access the channel. Thus the outputs of the designed DNN are joint channel access and MCS selection.

\subsection{Labeling and Training Procedure}

Before using supervised learning to train the DNN of the DL-MAC, we need to assign a label to each data. At time $t$, the label to the input data is the optimal result of the channel access and MCS selection adopted in the next packet transmission time, e.g., from time $t+1$ to $t+120$. Thus, the label to each data is the MCS selected from table~\ref{tab_MCS} to achieve the highest transmission rate below a target PER.  

We denote the MCS indexed by $i$ in table~\ref{tab_MCS} by $MCS_i$. It is known that PER depends on the receiving Signal-to-Interference-plus-Noise Ratio (SINR) and the adopted $MCS_i$, i.e., we can write $PER=f(SINR, MCS_i)$. And when SINR is fixed, PER is a monotonically increasing function of $MCS_i$. Therefore, for each $MCS_i$, subject to achieving PER below a target $T_p$ ($PER \leq T_p$), the minimum SINR, denoted by $SINR_i^{\min}$, can be derived by finding the SINR that satisfying: ${f\left( {SINR_i^{\min},MC{S_i}} \right) = T_p}$. After that, for each $MCS_i$, we can determine its operational SINR range where $MCS_i$ could achieve the highest transmission rate under the target PER as: $\left[SINR_{i}^{\min}, SINR_{i+1}^{\min}\right)$. We summarize the derived  operational SINR range for each $MCS_i$ in Table~\ref{tab_MCS}, and the complete workflow for data labeling is given as follows:
\begin{enumerate}
	\item Calculate average RSSI: At each time $t$, we collect one data that is a vector including the RSSIs of the previous 360 mini-slots, i.e., $\boldsymbol{D}=(RSSI_{t-359}^\prime, ...,  RSSI_{t}^\prime)$. We then observe the RSSIs of the next packet transmission time, i.e., $(RSSI_{t+1}^\prime,...,RSSI_{t+120}^\prime)$, and calculate the average RSSI of them: $\overline {RSSI}= \frac{1}{{120}}\mathop \sum \nolimits_{l = 1}^{l = 120} RSSI_{t + l}^{\prime}$.

	\item Calculate SINR: We transfer the $\overline {RSSI} $ into the average SINR at the receiver: $SINR=P_r - \overline {RSSI}$, where $P_r$ denotes the receive power at the receiver, which can be readily estimated at the transmitter in practical systems. We assume it to be equal to -60dBm without loss of generality in our experiments.
    \item Lookup MCS from Table~\ref{tab_MCS}: According to the calculated SINR of the next packet transmission time and the operational SINR ranges given in Table~\ref{tab_MCS}, we assign the MCS as the label to the input data at time $t$, e.g., if SINR is 22 dB, the assigned label is $MCS_5$ for $\boldsymbol{D}$.
\end{enumerate}

After finishing the labeling operation, we train the DNN using the labeled RSSI data vectors by using the Adam stochastic gradient descent algorithm \cite{kingma2014adam} with the cross-entropy loss function.

\subsection{Practical Implementation}
After the DNN is trained, we deploy the trained DNN at the AI device to implement our DL-MAC, i.e., the joint channel access and MCS selection. 

At the beginning, DL-MAC starts with an idle state and listens to the channel. Then, the AI device collects one RSSI data at each mini-slot $t$ and puts it into the tail of the RSSI data queue. DL-MAC feeds the recent 360 RSSI data from the queue to the DNN and outputs a selected $MCS_i$. If $MCS_{-1}$ is selected (i.e., not to transmit), the AI device keeps listening to the channel and then collects one more RSSI data from the next mini-slot $t+1$. If the output of the DNN is to transmit, the AI device transmits the oldest packet from its transmit buffer using the selected $MCS_{i}$ ($i \ne - 1$), over the mini-slot time $t+1$ to $t+120$.

When transmitting, the AI device cannot receive any signals from the wireless channel due to the half-duplex constraint on radio hardware. Thus, we need to generate RSSI data in a handcrafted manner for this packet transmission time after the transmission is finished; and put the handcrafted RSSIs into the RSSI data queue, so that these handcrafted RSSIs can be fed to the DNN as input to predict channel access and MCS in the next packet time. Consider that $MCS_i$ is selected for transmitting the packet. If the packet is successfully transmitted, it is supposed that the interference from other devices is low and thus the RSSI data are generated uniformly over the range between the RSSI corresponding to $SINR^{min}_i$ and the RSSI corresponding to $SINR^{min}_8$. On the contrary, if the packet transmission fails, it is supposed that the interference from other devices is high and thus the RSSI data are generated uniformly over the range between the RSSI corresponding to $SINR^{min}_{-1}$ and $SINR^{min}_{i-1}$.

\begin{table}[!t]
\centering
\caption{\tiny{THE MCS TABLE AND THE SINR OPERATION RANGE CORRESPONDING TO THE MCS INDEX WITH PER $\leq$ 10\%.}}
\vspace{-3mm}
\resizebox{65mm}{12mm}{
\begin{tabular}{|c|c|c|c|c|}
\hline
\textbf{MCS Index} & \textbf{Modulation} & \textbf{\begin{tabular}[c]{@{}c@{}}Coding Rate\end{tabular}} & \textbf{\begin{tabular}[c]{@{}c@{}}Rate(Mbps)\end{tabular}} & \textbf{\begin{tabular}[c]{@{}c@{}}SINR(dB)\end{tabular}} \\ \hline
-1 & (IDLE) & / & / & \textless 9 \\ \hline
0 & BPSK & 1/2 & 6.5 & {[}9, 10) \\ \hline
1 & QPSK & 1/2 & 13 & {[}10, 12) \\ \hline
2 & QPSK & 3/4 & 19.5 & {[}12, 15) \\ \hline
3 & 16-QAM & 1/2 & 26 & {[}15, 18) \\ \hline
4 & 16-QAM & 3/4 & 39 & {[}18, 21) \\ \hline
5 & 64-QAM & 2/3 & 52 & {[}21, 23) \\ \hline
6 & 64-QAM & 3/4 & 58.5 & {[}23, 24) \\ \hline
7 & 64-QAM & 5/6 & 65 & {[}24, 28) \\ \hline
8 & 256-QAM & 3/4 & 78 & $\geq $28 \\ \hline
\end{tabular}}
\label{tab_MCS}
\vspace{-5mm}
\end{table}

\section{EXPERIMENT RESULTS}
This section evaluates the performance of the DL-MAC via  simulations. We first introduce the data sampling environment and experimental setup, and then we discuss the simulation results of our DL-MAC and the benchmarks.

\subsection{Sampling Environment and Setup}
We conducted data sampling at two different indoor scenarios: our lab at Shenzhen University and the departure hall of Shenzhen Baoan International Airport. In the lab, we set up 12 Wi-Fi APs and 8 Bluetooth devices when conducting data sampling. The locations of the APs and Bluetooth devices are randomly deployed. Our data sampling hardware, was placed at the center position. In the airport departure hall, we sampled RSSIs at 5 different locations, i.e., check-in counters B, D, E, G, and the VIP lounge. The layouts of the two indoor scenarios and testbed can be found on our data set website \cite{postman511RSSIdataset}.

In the simulation experiments, the payload size of each packet is fixed to 1500 bytes, and packets arrive at transmitting buffer according to the Poisson distribution with a rate of $\lambda=0.18$. The interference threshold is set to -75dBm for CSMA/CA. For each sampling data that lasts for a period of time on a Wi-Fi channel, we use the first 100 seconds of the data to train the DNN, and use the following 20 seconds of the data to verify the performance of the trained DNN.

\subsection{Benchmark}
For benchmarking, we employ CSMA/CA and some commonly used rate adaptation algorithms such as I\textsc{wl} \cite{grunblatt2019simulation}, ARF \cite{kamerman1997wavelan} as the benchmarks to our DL-MAC. Moreover, we develop a global optimal method for the joint channel access and rate adaptation, referred to as the global optimal (GOPT) MAC. Furthermore, we employ the DL-based channel access and DL-based MCS selection benchmarks to verify the performance gain from the join design against the individual DL-based approaches. We itemize these benchmarks for clarity as follows:
\begin{itemize}
	\item GOPT: the global optimal scheme. For GOPT, the packet is not immediately transmitted but it can wait for some mini-slots before transmitting, if it finds that it can transmit at a higher rate from a latter mini-slot. Therefore, it can complete the packet transmission earliest even it needs to wait for some mini-slots.
	
	\item CSMA/CA+I\textsc{wl}: employing CSMA/CA for channel access and I\textsc{wl} algorithm to carry out MCS selection.
	
	\item CSMA/CA+ARF: employing CSMA/CA for channel access and ARF algorithm to carry out MCS selection.
	
	\item CSMA/CA+DL-MCS: employing CSMA/CA for channel access and DL-based algorithm for MCS selection.

	\item DL-CA+I\textsc{wl}: employing DL-based algorithm for channel access and I\textsc{wl} algorithm for MCS selection.
\end{itemize}

\subsection{Throughput Performance}
We use throughput as the performance metric to evaluate and compare the performance of our proposed DL-MAC with other benchmarks. We measure the throughput every 2 seconds.

\begin{figure}[!htb]
\vspace{-3mm}
\centering
\subfigure[]{
\includegraphics[width=3.5in]{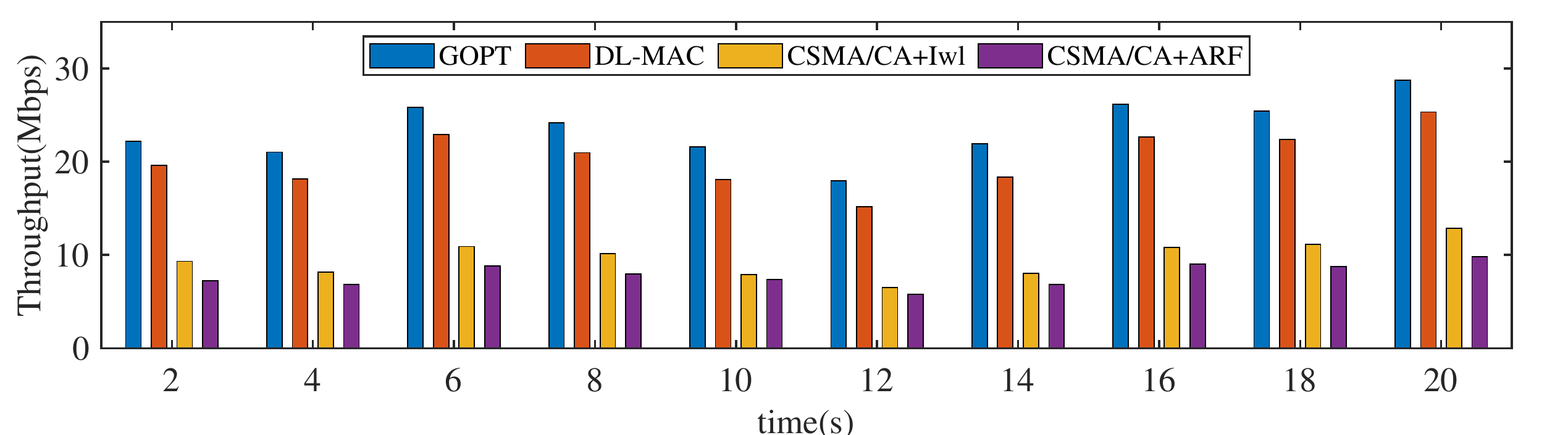}
\label{fig_dlmac_lab_6}
}
\subfigure[]{
\includegraphics[width=3.5in]{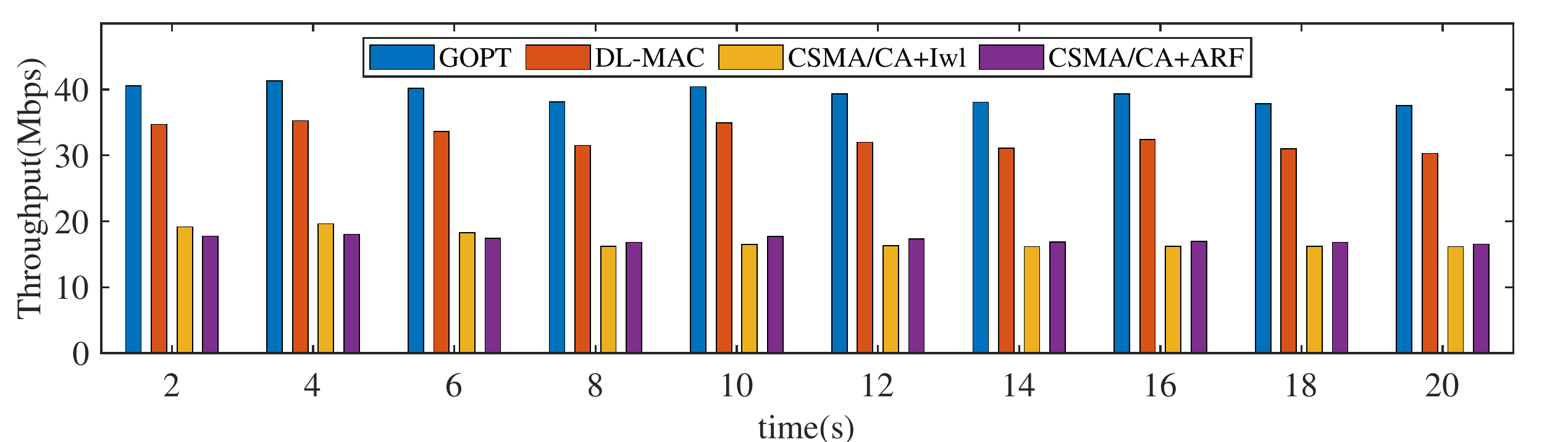}
\label{fig_dlmac_airport_1}
}
\vspace{-3mm}
\caption{Average throughputs of the DL-MAC for 10 independent experiments: (a) on channel 6 at the Shenzhen University lab, (b) on channel 1 at check-in counter G of the Shenzhen Baoan International Airport departure hall.}
\label{fig_DL_MAC}
\vspace{-3mm}
\end{figure}	 

 i) We first present the throughput performances of the two different scenarios, i.e., our lab in Shenzhen University and the departure hall of Shenzhen Baoan International Airport, in Fig.~\ref{fig_dlmac_lab_6} and Fig.~\ref{fig_dlmac_airport_1}, respectively. It can been seen that the performances of our proposed DL-MAC can approach that of the GOPT MAC, and outperform that of CSMA/CA with traditional rate adaptation algorithms. Specifically, in the environment of our lab, the DL-MAC can achieve 86.6\% of the performance of the GOPT MAC, and is more than around twice the performance of CSMA/CA and traditional rate adaptations. Similarly, in the scenario of the Shenzhen Baoan International Airport departure hall, the performance of the DL-MAC can achieve about 83.2\% of that of the GOPT MAC, and is higher than that of CSMA/CA with traditional rate adaptation schemes by about 91.3\%. It can be indicated that our DL-MAC presents significant advantages in further improving the spectrum efficiency of Wi-Fi networks.
 
\begin{figure}[!t]
\centering
\subfigure[]{
\includegraphics[width=3.5in]{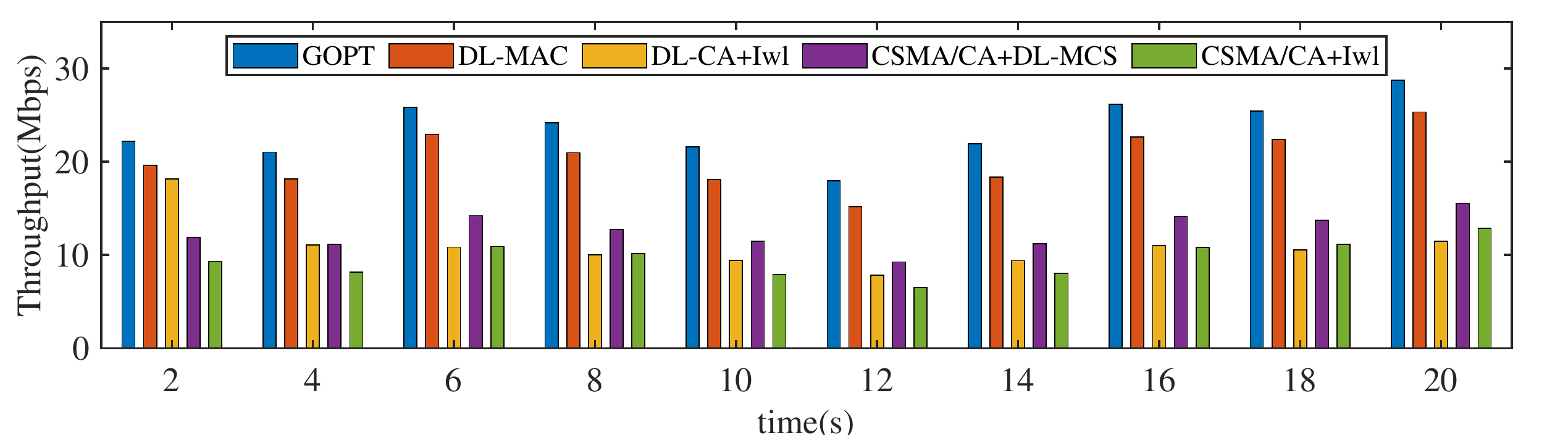}
\label{fig_dlmac_lab_6_vs}
}
\subfigure[]{
\includegraphics[width=3.5in]{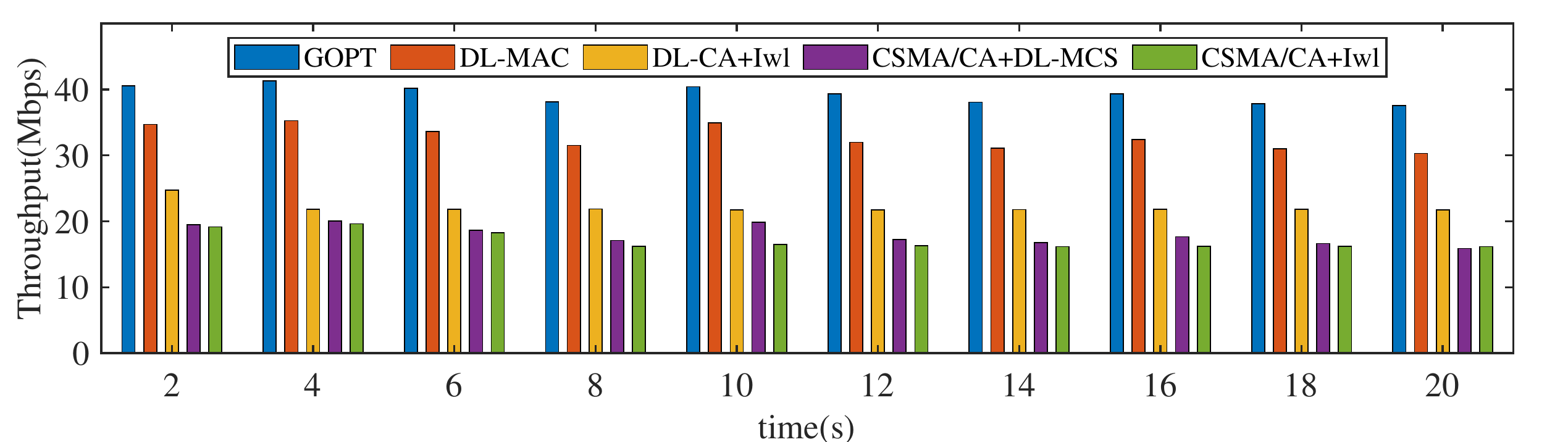}
\label{fig_dlmac_airport_1_vs}
}
\vspace{-3mm}
\caption{Comparison of performance gain of joint design and individual DL-based channel access and MCS selection, averaged of 10 independent experiments: (a) on channel 6 at the Shenzhen University lab, (b) on channel 1 at check-in counter G of the Shenzhen Baoan International Airport departure hall.}
\label{fig_DL_MAC_vs}
\vspace{-4mm}
\end{figure}	 
  
 ii) We next present the experimental results of the individual design of the DL-based channel access and the DL-based rate adaptation, to analyze the respective gains yield by channel access and rate adaptation in the joint design. Benchmarked to the individual design of the DL-based channel access and the DL-based MCS selection, our DL-MAC still shows an outstanding capability. In Fig.~\ref{fig_dlmac_lab_6_vs}, the average throughput of the DL-MAC has approximate 85.6\% gain over that of the DL-based channel access along with the I\textsc{wl} algorithm; while the average throughput of the DL-MAC is about 62\% higher than that of the combined scheme of the DL-based MCS selection along with CSMA/CA for channel access. In the joint design, MCS selection achieves a larger gain than channel access. Thus, we can infer that MCS selection is of vital importance to achieve a high throughput in a gentle channel condition. However, in Fig.~\ref{fig_dlmac_airport_1_vs}, in the scenario of the Shenzhen Baoan International Airport departure hall, the average throughput of the DL-MAC is about 82.3\% higher than the DL-based MCS selection using CSMA/CA for channel access, and approximate 47.9\% better than the DL-based channel access with the I\textsc{wl} algorithm. Contrary to the above result, the performance gain obtained from channel access is higher than that of obtained from MCS selection. The reason is that under severe channel conditions, channel access has more significant impact than MCS selection on the throughput improvement.
 
\begin{figure}[!ht]
\centering
\subfigure[]{
\includegraphics[width=1.65in]{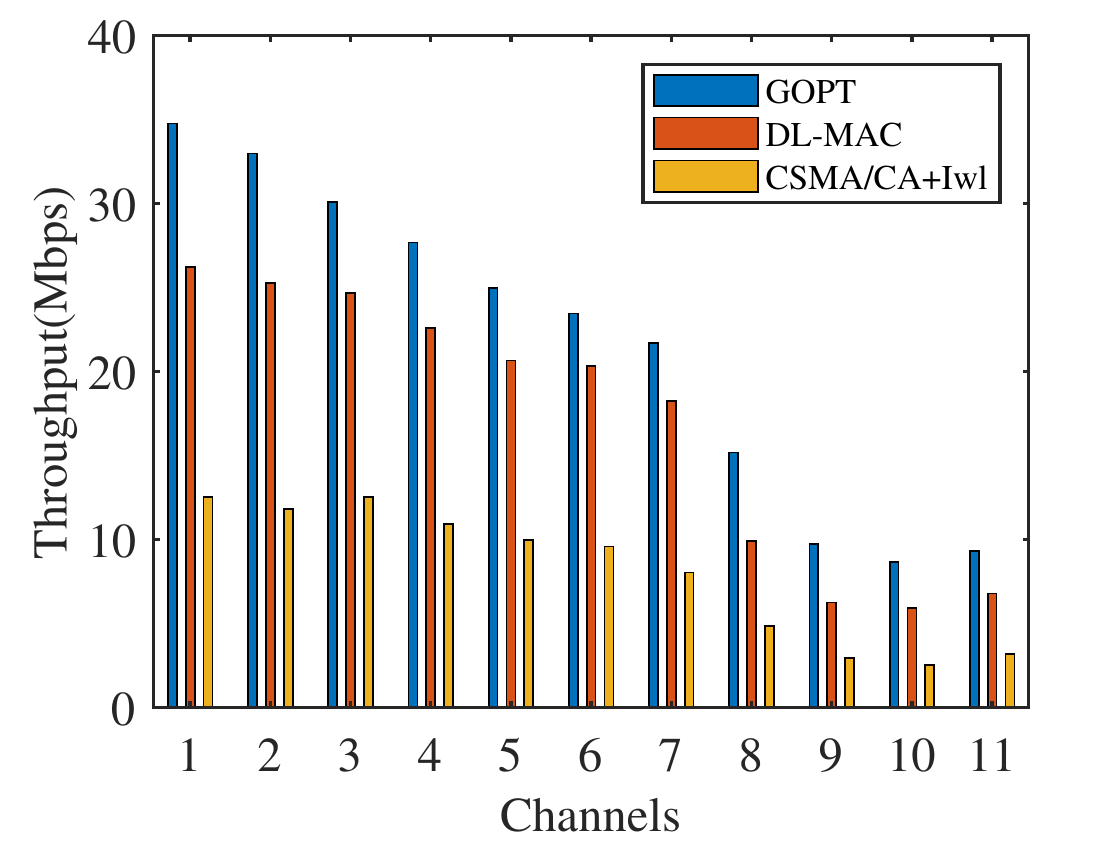}
\label{fig_g_ch11_lab_6}
}
\hspace{-3mm}
\subfigure[]{
\includegraphics[width=1.65in]{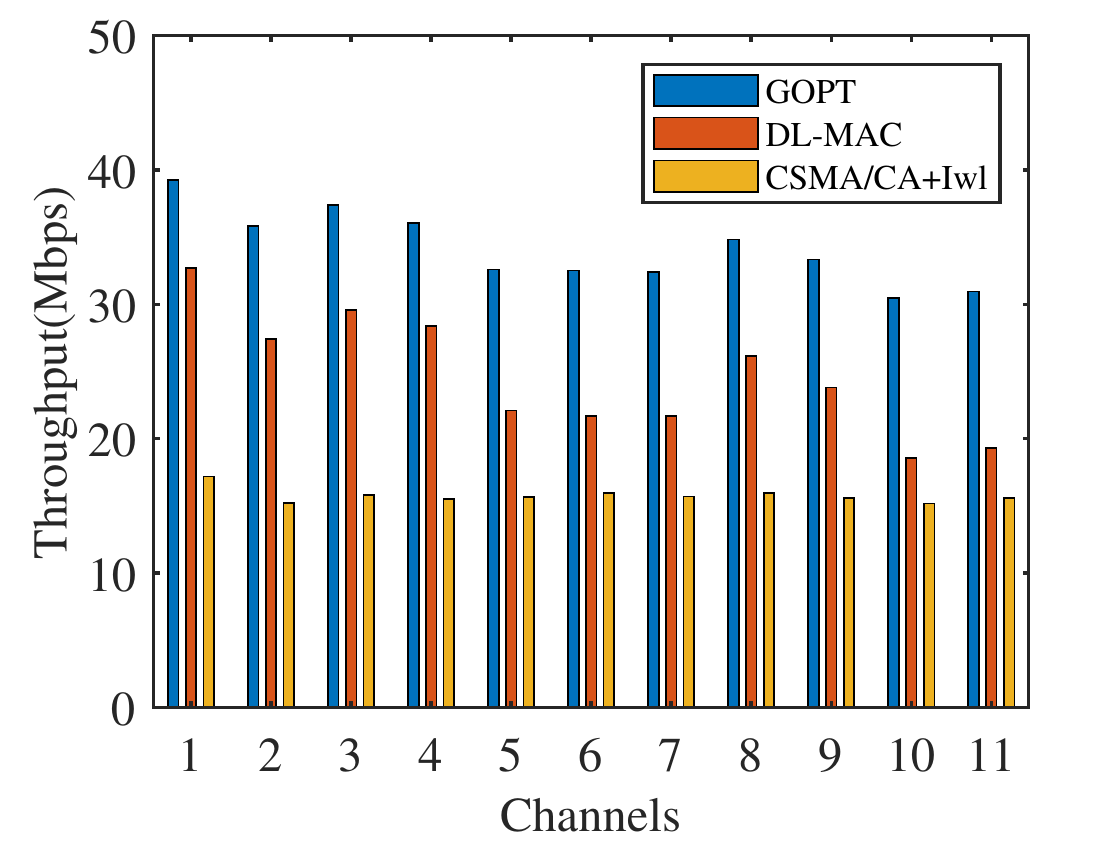}
\label{fig_g_ch11_airport_1}
}
\vspace{-4mm}
\caption{Throughput of DL-MAC over all channels, averaged of 10 independently experimental results: (a) at the Shenzhen University lab; (b) at check-in counter G of the Shenzhen Baoan International Airport departure hall.}
\label{fig_g_ch11}
\vspace{-3mm}
\end{figure}

\subsection{Generalization}
How well does our proposed DL-MAC perform on the channels that have never been used during training? This question is mainly about the generalization capability of the DL-MAC. We take DNN models that are seperately trained on channel 6 at the Shenzhen University lab and channel 1 at check-in counter G of the Shenzhen Baoan International Airport departure hall, to respectively evaluate the performances of the DL-MAC over 20 seconds over all channels. From Fig.~\ref{fig_g_ch11}, the generalization capability of the DL-MAC seems to be robust over all channels and outperforms the benchmark of using the CSMA/CA for channel access and traditional rate adaptation algorithm.

\begin{figure}[!t]
\centering
\subfigure[]{
\includegraphics[width=3.45in]{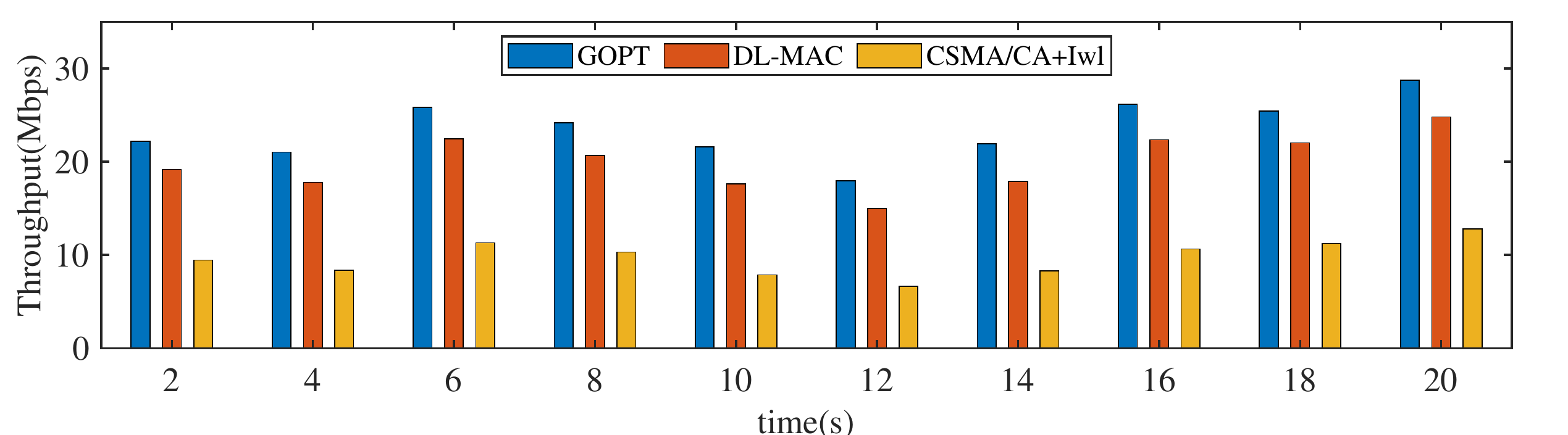}
\label{fig_lab_6_fusion}
}
\quad
\subfigure[]{
\includegraphics[width=3.45in]{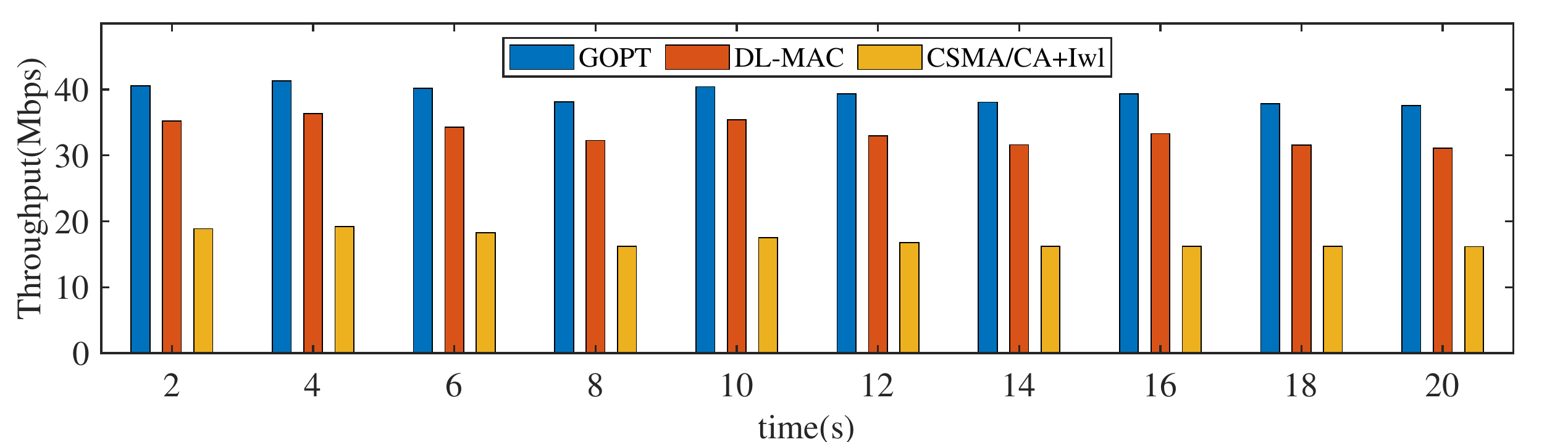}
\label{fig_airport_G_1_fusion}
}
\vspace{-3mm}
\caption{Throughput of the DL-MAC by fused data, averaged of 10 independent experiments: (a) on channel 6 at the Shenzhen University lab, (b) on channel 1 at check-in counter G of the Shenzhen Baoan International Airport departure hall.}
	\label{fig_fusion}
	\vspace{-3mm}
\end{figure}

Note that all the above experiments evaluate the performances of DNN for DL-MAC using the data from the same scenarios. However, we want to verify that the trained DNN for DL-MAC can also work for different scenarios, i.e., one trained DNN can work for both of the Shenzhen University and the Shenzhen International Airport scenarios. Considering the variability of data features sampled in different situations, we fuse the two data set into one data set to train DNN. Specifically, we use the 100 seconds of the sampled data on channel 6 at out lab and channel 1 at the airport for data fusion. Then we use the fused data to train the DNN. After that, we verify the performance of the DL-MAC. Fig.~\ref{fig_fusion} presents the throughput results of the DL-MAC whose DNN model is trained using the fused data. We can see that the DL-MAC still shows a high throughput and approaches the GOPT MAC, with outstanding performance at both our lab and the airport scenarios.
\vspace{-2mm}

\section{Conclusion}
In this work, we have investigated an intelligent DL-MAC protocol that can achieve joint channel access and MCS selection. Our DL-MAC is designed using DNN under supervised learning. DL-MAC aims at improving the spectrum efficiency of Wi-Fi networks on the 2.4GHz band. Our DL-MAC exploits the real wireless data sampled on the 2.4GHz band in  actual wireless environments. Then we experimentally verify the performance of DL-MAC, and the experimental results show that the DL-MAC  can achieve more efficient channel access with higher transmission rates compared to the CSMA/CA channel access and traditional rate adaptation schemes. Since our DL-MAC is developed from real sampled data, it is very practical and can work in real wireless networks.

\normalem
\bibliographystyle{IEEEtran}
\bibliography{refs}

\end{document}